\documentclass[%
 reprint,
showpacs,
 amsmath,amssymb,
 prl
]{revtex4-1}

\usepackage{graphicx}
\usepackage{dcolumn}
\usepackage{bm}
\usepackage{float}
\usepackage{xcolor}
\usepackage{amsmath}
\usepackage{amssymb}
\usepackage{amsfonts}

\newcommand{\f}{\frac}
\newcommand{\p}{\partial}
\newcommand{\m}{\mathbf}

\begin{document}

\preprint{APS/123-QED}

\title{Non-modal stability analysis and transient growth in a magnetized Vlasov plasma}

\author{Valeria Ratushnaya}

\author{Ravi Samtaney}

\affiliation{
 Mechanical Engineering, Physical Sciences and Engineering Division,  King Abdullah University of Science and Technology, 4700 KAUST, Thuwal 23955-6900, Kingdom of Saudi Arabia
\\
}


\begin{abstract}
Collisionless plasmas, such as those encountered in tokamaks,  exhibit a rich variety of instabilities. The physical origin, triggering mechanisms and fundamental understanding of many plasma instabilities, however, are still open problems. We investigate the stability properties of a collisionless Vlasov plasma in
a stationary homogeneous magnetic field. We narrow the scope of our investigation to the case of Maxwellian plasma. For the first time using a fully kinetic approach we show the emergence of the local instability, a transient growth, followed by classical Landau damping in a stable magnetized plasma. We show that the linearized Vlasov operator is non-normal leading to the algebraic growth of the perturbations using non-modal stability theory. The typical time scales of the obtained instabilities are of the order of several plasma periods. The first-order distribution function and the corresponding electric field are calculated and the dependence on the magnetic field and perturbation parameters is studied. Our results offer a new scenario of the emergence and development of plasma instabilities on the kinetic scale.
\end{abstract}


\maketitle

{\em Introduction}.---Stability theory, with its origins in mechanics, has had a long history with works in the 18-19th century and contributions  by mathematicians and physicists \cite{Lagrange1}-\cite{Leine}.
The main strategy is a consideration of small amplitude disturbances imposed on the base state, and linearization of the equations. Assuming the exponential time dependence of the perturbations, reduces the initial Cauchy problem to an eigenvalue problem of finding a spectrum of the governing operator. Then a dispersion relation (frequency as a function of the wave-number) is derived and depending on the sign of the real part of the eigenvalues one may deduce the stability characteristics of the system under investigation. This analysis essentially determines the stability as $t\rightarrow\infty$.
The further evolution of the stability theory, also known as modal stability theory, was successful in explaining many physical phenomena in fluid mechanics, plasma physics etc.
However, there were still a number of discrepancies between theory and experimental observations, which the classical stability theory failed to resolve, even by extending the existing classical theory to nonlinear orders. Problems of finite time-scale stability in hydrodynamics are the typical examples among such unresolved problems.
One of the most challenging puzzles was the discrepancy between the calculated Reynolds numbers and the experimentally observed critical Reynolds numbers in wall-bounded shear flows (see, e.g., Ref.~\cite{Tref1993}).

In the 1990s, classical stability theory underwent serious reconsideration resulting in the emergence of the non-modal stability theory, with the realization that the stability of a system must be understood in a broader sense. Arguably, hydrodynamic stability theory was the first branch of science that experienced a noticeable change in this regard. Farell {\em et al.} showed that a system can exhibit an instability or, so called transient growth, on finite times scales having all the eigenvalues in the lower half-plane, i.e. stable modes \cite{Far1982}-\cite{Far1996}. It was discovered that the reason of transient growth is the non-orthogonality of the eigenfunctions of the governing linear operator, i.e. a non-normality of the linear operator. The linearized Navier-Stokes operator is an example of a non-normal operator, and many fluid systems can experience a period of transient growth during which the perturbations increase in magnitude~\cite{SchmidBook, Schmid2007}. Typical examples are the parallel viscous shear flows such as plane Poiseuille and Couette flows (see, e.g., Refs.~\cite{Reddy1993, ReddyJFM1993}). The stability of such flows was reconsidered with no assumptions about the form of the disturbances and the discrepancy between the calculated and observed Reynolds numbers was eliminated \cite{Butler1992, Tref2005}. The information about the stability on finite time-scales missing in the classical stability theory was recovered in the framework of non-modal approach \cite{Tref1993}-\cite{Tref2005}.

It is natural to question whether plasma systems such as tokamaks can also exhibit transient growth or short-time scale instabilities due to the non-normality of the governing operators. Due to infrequent or no collisions, a fluid treatment of plasma (e.g. using magnetohydrodynamics) is generally considered inadequate, and the most appropriate framework is kinetic theory. Recall that early attempts by Vlasov to examine the stability properties of the kinetic equation for collisionless plasma systems were somewhat flawed \cite{Vlasov1938,Vlasov1945}. This was realized and improved by Landau in Ref.~\cite{Landau1946}, where the Landau damping phenomenon was discovered, predicting stability of the Maxwellian collisionless plasma as $t\rightarrow\infty$.
In follow-on works on plasma stability (see, e.g., Refs.~\cite{Kampen, Bernstein}), the classical stability theory was used. Recently, Podesta addressed the issue of transient growth for the case of one-dimensional collisionless plasma and showed that, indeed, one may construct solutions of the linearized problem leading to the emergence of the short time-scale instabilities \cite{Podesta2010}.
In order to examine transient growth of instabilities for a magnetized plasma such as that encountered in tokamaks, it is imperative to consider the presence of a magnetic field. One such attempt to include a magnetic field was addressed  in Ref.~\cite{Campo2009} in the context of a  fluid model (due to presumed complications using kinetic theory), and the authors acknowledged that a fully kinetic approach based on Vlasov equation must be used in order to draw physically relevant conclusions regarding stability of a magnetized plasma. It is precisely the goal of this work to examine whether transient growth occurs in a collisionless plasma governed by the kinetic Vlasov equation in the presence of an external magnetic field. We turn our attention to the stability analysis of such a system next, on a purely kinetic level, and further address the issue of non-normality of the governing linearized stability operator.

{\em Vlasov magnetized plasma: formulation of the model and basic equations}.---We begin our discussion with the kinetic Vlasov equation for the distribution function $f\left(\m{r},\m{v},t\right)$, \cite{Vlasov1938} :
\begin{equation}\label{VE}
\f{\p f}{\p t}+\left(\m{v}\cdot\nabla\right)f+\f{q}{m}\left(\m{E}+\m{v}\times\m{B}\right)\cdot\f{\p f}{\p\m{v}}=0,
\end{equation}
where $q, m,\m{B}\left(\m{r},t\right)$ and $\m{E}\left(\m{r},t\right)$ are the electric charge, mass, magnetic field and electric field, respectively.
Additionally, these quantities are related self-consistently to each other by the Maxwell equations of electrodynamics. Here, we consider a stationary homogeneous external magnetic field, so that the electric field can be determined from an electrostatic potential. Following the tenets of classical linear stability analysis, we impose a small arbitrary perturbation $f_{1}$ on the equilibrium state $f_{0}$:
\begin{equation}\label{pert}
f\left(\m{r},\m{v},t\right)=f_{0}\left(v\right)+f_{1}\left(\m{r},\m{v},t\right),\quad f_{1}\ll f_{0}.
\end{equation}
Note we do not specify any particular functional form of the perturbation.
As mentioned above, the non-modal approach and full stability picture can be obtained only when the disturbance is considered in its most general form. The only assumption we make is the smallness of the perturbation compared with the equilibrium state of the plasma. Here we make the somewhat obvious choice of a Maxwellian equilibrium. Although the magnetic field introduces anisotropy resulting in different temperatures in directions perpendicular and parallel to it, without loss of generality, an isotropic equilibrium is assumed given by
\begin{equation}\label{equil}
f_{0}\left(v\right)=\f{n_{0}}{\left(2\pi\right)^{3/2}v^{3}_{T}}\exp\left(-\f{v^{2}}{2v^{2}_{T}}\right),
\end{equation}
where $v_{T}=\sqrt{K_{B}T/m}$ is the thermal speed, $T$ is the temperature and $K_{B}$ is the Boltzmann constant. As usual, this distribution function is normalized to the particle density $n_{0}$ to give $\int f_{0}\left(v\right)\m{dv}=n_{0}$.

{\em Linear stability analysis: non-normality of the linear operator}.---We obtain the linearized equation by substituting Eq.~\eqref{pert}, into the Vlasov equation, Eq.~\eqref{VE}, and retaining terms up to first order only:
\begin{equation}
\f{\p f_{1}}{\p t}+\left(\m{v}\cdot\nabla\right)f_{1}+
\f{q}{m}\left[\left(\m{v}\times\m{B}\right)\cdot\nabla_{\m{v}}f_{1}+
\left(\m{E_{1}}\cdot\nabla_{\m{v}}\right)f_{0}\right]=0.
\label{LVE}
\end{equation}
A traditional way to solve this type of problem is to use the method of characteristics followed by the Fourier and Laplace transformations \cite{Landau1946,Bernstein,Nicholson1983}. It is more convenient for us first to Fourier transform Eq.~\eqref{LVE} leading to
 \begin{equation}
\begin{split}
\f{\p f_{1,\m{k}}}{\p t}+i\left(\m{k}\cdot\m{v}\right)f_{1,\m{k}}+\f{q}{m}\left(\m{v}\times\m{B}\right)\cdot\nabla_{\m{v}}f_{1,\m{k}}\\
+\f{q}{m}\left(\m{E_{1,\m{k}}}\cdot\nabla_{\m{v}}\right)f_{0}=0.\label{VEF}
\end{split}
\end{equation}
We henceforth omit subscript $1,\m{k}$ so the notation $f, \m{E}$ refers to the Fourier transformed first-order distribution function and the electric field, respectively.
The dynamics of the electric field is determined by the Fourier-transformed linearized Maxwell equation for the curl of the magnetic field:
\begin{equation}
\f{\p\m{E}}{\p t}=-\f{q}{\epsilon_{0}}\int\m{v}f\m{dv}.\label{P}
\end{equation}
Rewriting Eqs.~\eqref{VEF} and \eqref{P} in the form of a linearized dynamical system,
\begin{equation}\label{linA}
\f{\p\m{X}}{\p t}=\hat{A}\m{X},\,\,\text{where}\,\,\m{X}= \left( \begin{array}{c}
f \\
\m{E}
\end{array} \right)\!, \hat{A} = \left( \begin{array}{cc}
a_{11} & a_{12}\\
a_{21} & a_{22}\end{array} \right),
\end{equation}
where $\hat{A}$ a linear operator of the magnetized Vlasov plasma, or Vlasov operator, with the following elements:
\begin{eqnarray}
&&a_{11}=-i\left(\m{k}\cdot\m{v}\right)-\f{q}{m}\left(\m{v}\times\m{B}\right)\cdot\f{\p}{\p\m{v}},\nonumber\\
&&a_{12}=-\f{q}{m}\f{\p f_{0}}{\p\m{v}}\cdot,\quad a_{21}=-\f{q}{\epsilon_{0}}\int\m{v}\m{dv}\cdot,\quad a_{22}=0,\label{a22}
\end{eqnarray}
where ``$(\cdot)$'' denotes the inner product. A necessary condition for a system governed by Eq.~\eqref{linA} to exhibit transient growth, even if the eigenvalues of $\hat{A}$ indicate stability, is the non-normality of the operator $\hat{A}$, i.e., $\hat{A}\hat{A}^{+}\neq \hat{A}^{+}\hat{A}$, where $\hat{A}^{+}$ is the adjoint of $\hat{A}$. Straightforward algebra shows that the Vlasov operator given by Eq.~\eqref{a22} is indeed non-normal. A consequence of this non-normality, as shown later, is the emergence of the growth of the perturbations on short time scales corresponding to the plasma period.

\begin{figure}[h!]
\centering
\vspace{-0.4cm}
\includegraphics[width=0.6\linewidth]{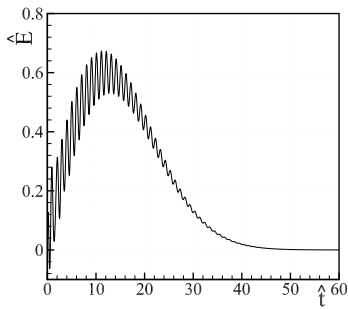}
\caption{Transient growth of the electric field for $\hat{k}=0.25$, $\alpha=0.05$ and $\hat{\omega}=2\pi$.}
\label{Eoft}
\vspace{-0.4cm}
\end{figure}

{\em Non-modal stability approach and transient growth}.---We employ the method of integration along the particles unperturbed trajectories, which is one of the standard methods to solve Eq.~\eqref{VEF}, (the reader may consult, for example, Ref.~\cite{Bellan2006}). We introduce new variables $\left(t',\m{v'}\right)$, such that
\begin{equation}\label{unperttraj}
\f{d\m{v'}}{dt'}=\f{q}{m}\m{v}\times\m{B}, \quad \text{with} \quad \m{v'}\left(t=t'\right)=\m{v},
\end{equation}
meaning that for all times $t<t'$ the dynamics of charges is described simply by Lorentz equations.
This implies that Eq.~\eqref{VEF} can be rewritten as a non-homogeneous ordinary differential equation:
\begin{equation}\label{ode}
\f{df}{dt'}+i\left(\m{k}\cdot\m{v'}\right)f+\f{q}{m}\left(\m{E}\cdot\nabla_{\m{v'}}\right)f_{0}=0.
\end{equation}
For simplicity we assume the magnetic field $\m{B}$ in the Z-direction, and the wave vector $\m{k}$ in the XOZ-plane. Under the electrostatic approximation, it is evident that in our new reference frame the electric field is parallel to the wave vector so that $\f{E}{k}\m{k}\cdot\m{v'}=\m{E}\cdot\m{v'}$.
Integrating Eq.~\eqref{unperttraj} gives the unperturbed trajectory
\begin{eqnarray}
v'_{x}&=&v_{x}\cos\omega\left(t-t'\right)-v_{y}\sin\omega\left(t-t'\right),\nonumber \\
v'_{y}&=&v_{x}\sin\omega\left(t-t'\right)+v_{y}\cos\omega\left(t-t'\right),\nonumber \\
v'_{z}&=&v_{z},
\end{eqnarray}
where $\omega=qB_{0}/m$ is the cyclotron frequency and $B_{0}$ is the magnitude of the magnetic field $\m{B}$.
These trajectories are taken into account in Eq.~\eqref{ode} to obtain the first-order distribution function
\begin{align}\label{f11}
&f\!=\!f\!\left(\m{k},\m{v},0\right)\exp
\left\{-i\left(A_{t}v_{x}-B_{t}v_{y}+k_{z}v_{z}t\right)\right\}\nonumber \\
&\!\!+\!\f{q}{mk}\!\int_{0}^{t}\!\!\f{d\tau}{v}\f{df_{0}}{dv}E\left(v_{x}\left(k_{x}-\omega B_{t-\tau}\right)-v_{y}\omega A_{t-\tau}+k_{z}v_{z}\right)\nonumber\\
&\times\exp\left\{-i\left(v_{x}A_{t-\tau}-v_{y}B_{t-\tau}+k_{z}v_{z}\left(t-\tau\right)\right)\right\},
\end{align}
describing the perturbations propagating in a collisionless magnetized Vlasov plasma, where
\begin{equation}
A_{t}=\f{k_{x}}{\omega}\sin\omega t,\quad\text{and}\quad B_{t}=\f{k_{x}}{\omega}\left(1-\cos\omega t\right).
\end{equation}
To demonstrate transient algebraic growth of perturbations followed by late-time Landau damping \cite{Landau1946}, we focus on the temporal evolution of the electric field, namely, we use the Fourier-transformed Maxwell equation for the divergence of the electric field:
\begin{equation}
E\left(\m{k},t\right)=-\f{q}{i\epsilon_{0}k}\int f\left(\m{k},\m{v},t\right)\m{dv},
\end{equation}
where we have taken into account the fact that the electric field is parallel to the direction of propagation.
Substitution of the distribution function, Eq.~\eqref{f11}, into the previous equation results in following Volterra equation of the second kind,
\begin{equation}\label{inteq}
E\left(\m{k},t\right)=J\left(\m{k},t\right)+\int_{0}^{t}K\left(\m{k},t-\tau\right)Ed\tau,
\end{equation}
where
\begin{align}\label{J}
&J\left(\m{k},t\right)=-\f{q}{i\epsilon_{0}k}\int_{-\infty}^{+\infty}f\left(\m{k},\m{v},0\right)\\
&\times\exp\left\{-i\left(A_{t}v_{x}-v_{y}B_{t}+k_{z}v_{z}t\right)\right\}\m{dv}\nonumber
\end{align}
and the integral kernel is
\begin{align}\label{K}
&K\left(\m{k},t-\tau\right)=-\f{q^{2}}{i\epsilon_{0}mk^{2}}\int_{-\infty}^{+\infty}\f{1}{v}\f{df_{0}}{dv}\\
&\times\left(v_{x}(k_{x}-\omega B_{t-\tau}) -v_{y}\omega A_{t-\tau}+k_{z}v_{z}\right)\nonumber\\
&\times\exp\left\{-i\left(v_{x}A_{t-\tau}-v_{y}B_{t-\tau}+k_{z}v_{z}\left(t-\tau\right)\right)\right\}\m{dv}.\nonumber
\end{align}
Note that the function $J\left(\m{k},t\right)$ depends on the initial perturbation $f\left(\m{k},\m{v},0\right)$ while the kernel $K\left(\m{k},t-\tau\right)$ depends on the velocity derivative of the initial equilibrium function $f_0$. The freedom in $f\left(\m{k},\m{v},0\right)$ choice is limited only by the assumption of the smallness of the perturbations, Eq.~\eqref{pert}. Here we limit our choice to odd functions of $\m{v}$ so that the total number density of particles, $n_0$, is unchanged, and consider one such possibility:
\begin{equation}\label{fkv0}
f\left(\m{k},\m{v},0\right)=\f{v_{T}}{\left(2\pi\right)^{3/2}v^{5}_{f}}\left(\m{C}\cdot\m{v}\right)
e^{-v^{2}/2v^{2}_{f}}\!,
\end{equation}
where in general $\m{C}=\left(C_{x}, C_{y}, C_{z}\right)$ and $v_{f}$ are some characteristic parameters of the initial data. Note that a similar choice was made by Podesta \cite{Podesta2010}. These parameters are related to the amplitude and the duration of the transient regime in plasma. Further simplifying, we consider an isotropic case when $\m{C}=C_{0}\m{e}$ and $\m{v}=\f{v}{\sqrt{3}}\m{e}$, where $C_{0}$ is some constant and $\m{e}=\left(1, 1, 1\right)$. We further also consider isotropic wave vectors, i.e. $k_{x}=k_{z}\equiv k_{0}$. Moreover, we introduce the following dimensionless quantities:
\begin{eqnarray}
&&\hat{t}=\f{t}{\tau_p}=\f{\omega_{p}t}{2\pi},\ \hat{\omega}=2\pi\f{\omega}{\omega_{p}},\
\alpha=\f{v_{f}}{v_{T}},\  \hat{k}=k_{0}\lambda_{D},
\end{eqnarray}
where $\omega_p=\sqrt{n_{0}q^{2}/m\epsilon_0}$ is the plasma frequency,  $\tau_p$ is plasma period and $\lambda_{D}$ is the Debye length. Given initial profile, Eq.~\eqref{fkv0}, and the Maxwellian equilibrium, Eq.~\eqref{equil},  we derive the following expressions for $K\left(\m{k},t-\tau\right)$ and $J\left(\m{k},t\right)$ as follows
\begin{align}\label{K1}
&K\left(\m{k},t-\tau\right)\equiv \f{1}{\tau_p}\hat{K}(\hat{\m{k}},\hat{t}-\hat{\tau})\nonumber\\
&=-\f{1}{\tau_p}\left(2\pi^2\right)
\left(\left(\hat{t}-\hat{\tau}\right)+\f{1}{\hat{\omega}}\sin\hat{\omega}\left(\hat{t}-\hat{\tau}\right)\right) \\
&\times
\exp\left\{-2\pi^2\hat{k}^{2}\left[\f{2}
{\hat{\omega}^{2}}\left(1-\cos\hat{\omega}\left(\hat{t}-\hat{\tau}\right)\right)+\left(\hat{t}-\hat{\tau}\right)^{2}\right]\right\},\nonumber
\end{align}
and
\begin{align}\label{J1}
&J\left(\m{k},t\right)\equiv\nonumber\left(\f{qv_{T}}{\epsilon_0}\right)\hat{J}\left(\hat{\m{k}},\hat{t}\right)\\
&=\left(\f{qv_{T}C_{0}}{\epsilon_0}\right)\f{1}{\sqrt{2}}
\left(\f{\sin\hat{\omega}\hat{t}}{\hat{\omega}}-\f{\left(1-\cos\hat{\omega} \hat{t}\right)}{\hat{\omega}}+\hat{t}\right)\\
&\times
\exp\left\{-2\pi^2\alpha^2\hat{k}^{2}\left[\f{2}
{\hat{\omega}^{2}}\left(1-\cos\hat{\omega}\hat{t}\right)+\hat{t}^{2}\right]\right\},\nonumber
\end{align}
where the prefactor $\left(\f{qv_{T}}{\epsilon_0}\right)$ in $J(\m{k},t)$ has units of the Fourier transformed electric field. The non-dimensional equation governing the electric field is then
\begin{equation}\label{inteqnondim}
\hat{E}\left(\hat{\m{k}},\hat{t}\right)=\hat{J}\left(\hat{\m{k}},\hat{t}\right)+\int_{0}^{\hat{t}}\hat{K}\left(\hat{\m{k}},\hat{t}-\hat{\tau}\right)\hat{E}d\hat{\tau}.
\end{equation}
Within the context of the linear stability analysis, the amplitude $C_0$ may be scaled out, or normalized to unity. Evidently, the functions $J$ and $K$ for small times grow algebraically and are dominated by exponential decay at later times. A standard way to solve Eq.~\eqref{inteqnondim} involves Laplace transforms. However, in our case due to complexity of Eqs.~\eqref{K1} and \eqref{J1}, we solve Eq.~\eqref{inteqnondim} directly by numerical means (several verification and convergence tests were conducted and are not reported here). We note that one may independently choose the strength of the external magnetic field $B_0$, the equilibrium number density $n_0$ and the temperature $T$. These choices fix the thermal speed $v_T$, the Debye length $\lambda_D$ and the frequencies $\omega$ and $\omega_p$. After non-dimensionalization, the three essential parameters are: $\hat{\omega}$, the ratio of the two frequencies, and essentially the only non-dimensional parameter relevant to the dynamics of the Vlasov equation; $\alpha$, the parameter governing the initial perturbation of the distribution function; and the wave number $\hat{k}$.  In Figure~\ref{Eoft} we plot the solution of Eq.~\eqref{inteqnondim} for $\hat{k}=0.25$, $\alpha=0.05$ and $\hat{\omega}=2\pi$ (this non-dimensional frequency is chosen to be relevant to tokamak parameters so that for a choice of $B_0=1T$, $n_0=10^{19}/m^3$ and $T=10^8K$,  we obtain $\hat{\omega}\approx 2\pi$).
As it can be seen, there is a transient growth of the electric field over $\sim10-15$ plasma periods followed by the classical Landau damping. Depending on the parameters $\hat{\omega}$, $\hat{k}$ and $\alpha$, the duration and the amplitude of the transient growth can vary, thus one may prolong or shrink the instability region. Obviously, this instability is limited and cannot be extended to infinitely large times and amplitudes. The smallness of the perturbations, Eq.~\eqref{pert}, leads to a constraint imposed on the initial parameters, which for the planar waves, for instance, reduces to $C_{0}\ll\alpha^{4}\exp\left[\left(1-\alpha^{2}\right)/2\right]$.

\begin{figure}[h!]
\centering

\vspace{-0.4cm}

  \includegraphics[width=0.6\linewidth]{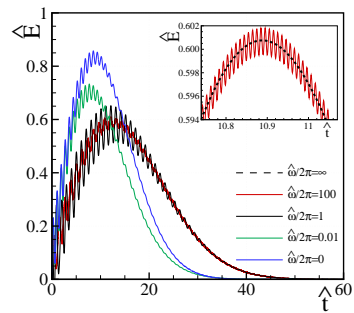}
  \caption{Dependence of $\hat{E}$ on the magnetic field (or non-dimensional parameter $\hat{\omega}$) for $\hat{k}=0.25$, $\alpha=0.05$.}
  \label{EB}
\end{figure}

{\em Transient growth: regimes and limits}.---We now study the dependence of the transient growth on the magnetic field strength. In Figure~\ref{EB} we plot time history of the electric field for $\hat{k}=0.25$ and $\alpha=0.05$ for various values of the non-dimensional frequency $\hat{\omega}$ ranging from $0$ (zero magnetic field) to $\infty$ (infinitely strong magnetic field).  For all values of $\hat{\omega}$ we observe the transient growth of the electric field over $10-15$ plasma periods. The solution exhibits two frequencies: one corresponding to the plasma frequency $\omega_p$ (non-dimensionalization leads to a plasma period of unity) and the second frequency corresponding to the cyclotron frequency. The gyroperiod for the case $\hat{\omega}=2\pi 100$ is $0.01\tau_p$ and is evident in the inset in Figure~\ref{EB}. The limiting case of $\hat{\omega}=0$ solution, which can be compared qualitatively to that in Ref.~\cite{Podesta2010}, exhibits only the plasma oscillations. As $\hat{\omega}$ increases the peak electric field decreases and occurs at a slightly later time, and the oscillations decrease in amplitude until the limiting case of $\hat{\omega}\rightarrow\infty$, which is shown as the dashed black curve in the inset in Figure~\ref{EB}.  The electric field solution follows the same profile as $\hat{J}$ (not shown) with its maximum lagging slightly behind the maximum of $\hat{J}$. The leading order term in $\hat{J}$ for $\hat{\omega}=0$ (resp. $\hat{\omega}\rightarrow\infty$) is $t\exp(-2\pi^2\alpha^2\hat{k}^2\hat{t}^2)$ (resp. $t\exp(-4\pi^2\alpha^2\hat{k}^2\hat{t}^2)$), and the time when maximum $\hat{J}$ (and an estimate of when maximum of $\hat{E}$) occurs is $t_{max}|_{\hat{\omega}=0}\!\!=\!\!1/(2\pi\alpha\hat{k})$ and $t_{max}|_{\hat{\omega}\rightarrow\infty}\!\!=\!\!1/(\sqrt{2}2\pi\alpha\hat{k})$. Thus increasing the magnetic field from $0$ to very high values shifts the maximum of the transient growth and reduces its value only by the factor of $\sqrt{2}$.
\begin{figure}[h!]
\centering

\vspace{-0.4cm}

  \includegraphics[width=0.6\linewidth]{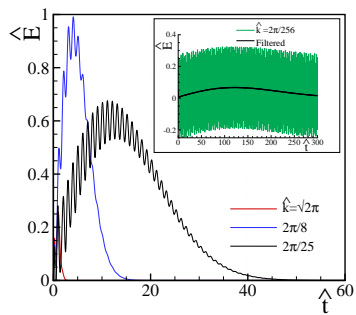}
  \caption{Dependence of $\hat{E}$ on the wave number $\hat{k}$ for $\alpha=0.05$, $\hat{\omega}=2\pi$.}
  \label{Ek}
\end{figure}
We now examine the dependence of $\hat{E}$ on the wave number $\hat{k}$. Perturbations of the wavelength shorter than Debye length are of no physical interest, thus we obtain a natural upper cut-off of the perturbations, i.e. $\hat{k}_{max}=\pi\sqrt{2}$, with the limit of $\hat{k}\rightarrow 0$ indicating very long wavelength perturbations.  Figure~\ref{Ek} shows the dependence of the transient growth on $\hat{k}$ for $\alpha=0.05$ and $\hat{\omega}=2\pi$. Perturbations on the scale of the Debye length die out very quickly on the order of a few plasma periods. As $\hat{k}$ decreases the peak $\hat{E}$ increases and then decreases with the time at which the maximum occurs shifting to the right. Also instabilities corresponding to the smaller $\hat{k}$ (longer wavelengths) have slower decay in time, which is in agreement with Landau theory, showing that the Landau damping is weaker for longer wavelengths. For very long wavelength perturbations (see inset in Figure~\ref{Ek}) we observe a very oscillatory solution with large amplitude. In order to see the tendency of the electric field, we filter the solution to remove these oscillations, revealing a small growth followed by the Landau damping over time scales of hundreds of plasma periods. Note that the limiting solution ($\hat{k}=0$) is simply oscillatory with no growth at all.

\begin{figure}[h!]
\centering

\vspace{-0.4cm}

  \includegraphics[width=0.6\linewidth]{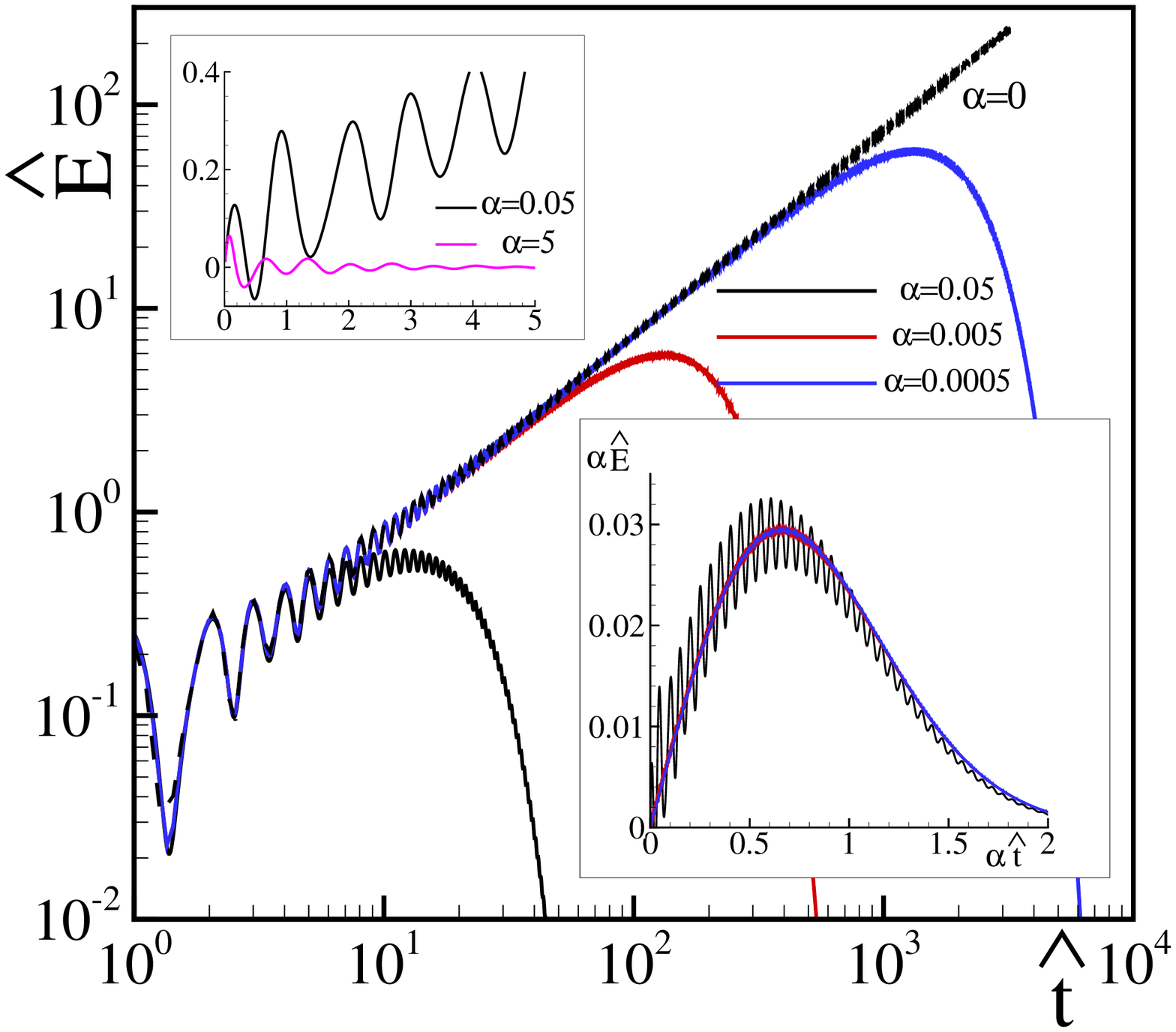}
  \caption{Dependence of $\hat{E}$ on $\alpha$ for $\hat{k}=0.25$, $\hat{\omega}=2\pi$.}
  \label{Ealpha}
\end{figure}

The dependence of $\hat{E}$ on $\alpha$ is shown in Figure~\ref{Ealpha}. From these plots we can draw several conclusions. The first one is that by varying $\alpha$ one can control the peak electric field and the duration of the transient growth. In particular, by setting the ratio between the characteristic velocities sufficiently small, one can enhance the amplitude of the transient growth and prolong the duration of this local instability (see Figure~\ref{Ealpha}). The second is that large $\alpha$ perturbations decay quickly in less than a couple of plasma periods (see top-left inset in Figure~\ref{Ealpha}). The asymptotic solution for $\alpha=0$ grows without limit in time and is plotted in Figure~\ref{Ealpha} for reference purposes. The first term in the series solution for $\alpha=0$ indicates that the time at which $\hat{E}$ attains its peak scales as $1/\alpha$ and the peak magnitude of $\hat{E}$ also scales as $1/\alpha$. This is borne out by numerical results in the bottom-right inset in Figure~\ref{Ealpha} wherein we plot the scaled electric field $\alpha \hat{E}$ vs. scaled time $\alpha\hat{t}$  for $\alpha=0.05,0.005,0.0005$, showing the universality of these solutions.

\begin{figure}[h!]
\centering

\vspace{-0.4cm}

  \includegraphics[width=0.6\linewidth]{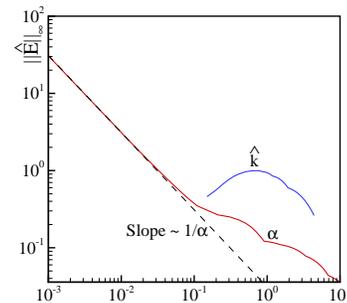}
  \caption{$||\hat{E}||_\infty$ as a function of $\hat{k}$ and $\alpha$ for $\hat{\omega}=2\pi$.}
  \label{Emaxalphak}
\end{figure}

We plot the maximum of $\hat{E}$, denoted by the $\infty-$norm $||\hat{E}||_\infty$, by varying $\alpha$ and $\hat{k}$ in Figure~\ref{Emaxalphak}. Maximal transient growth can be achieved when $\alpha$ approaches its minimal value, i.e. $\alpha=0$ and {\it vice versa}, the transient growth can be neglected, if $\alpha$ is infinitely large. The largest transient growth occurs for $\hat{k}\approx 0.7$. Another interesting observation concerns the dependence of the maximal transient growth for different wavelength perturbation regimes. As one can see in Figure~\ref{Emaxalphak}, the amplitude of the transient growth is not a monotonic function of $\hat{k}$. In other words, there is a specific perturbation of a specific wavelength leading to the highest value of the transient growth amplitude $||\hat{E}||_\infty$. In our case this value is $\hat{k}\approx 0.7$. All the perturbations of shorter or longer wavelengths lead to the lower amplitudes of the perturbations.
{\em Conclusions}.---The main contribution of the work presented here is that using a fully kinetic approach, we have investigated the stability properties of the magnetized Vlasov plasma in the magnetic field, and have clearly demonstrated that such a plasma exhibits transient growth. The obtained local instability emerges due to the non-normality of the governing operator and is followed by the classical Landau damping. The typical time scales of this transient growth is of the order of several plasma periods.
The existence of kinetic instabilities in magnetized plasma is of crucial importance in essentially collisionless plasma systems such as tokamaks.  Transient growth for a nominally stable plasma may be detrimental for plasma confinement, especially, as we have shown, that even in the limit of an infinite magnetic field the transient growth is barely suppressed. Therefore, it is important to identify such possible transient instabilities and study their physical properties and behavior. We have shown that depending on the initial conditions and perturbation parameters one can control the transient growth regime, i.e. its speed, duration and amplitude. The maximal transient growth is calculated and its behavior is studied as well. Our work is an initial step towards understanding the kinetic instabilities in tokamak physics. Although, we discuss the emergence of transient growth in a stationary homogeneous magnetic field, we conjecture that more complex magnetic field laboratory configurations will also exhibit such transient growth. Stability analysis of the magnetized plasma in more realistic magnetic field geometries is an ongoing project and will be a topic of our future work.

\end{document}